\documentclass[twocolumn,aps,prb,superscriptaddress,multicol,amsmath,amssymb,10pt]{revtex4-2}
\usepackage[utf8]{inputenc}
\usepackage[normalem]{ulem}
\usepackage{amsmath}
\usepackage{siunitx}
\usepackage{graphicx}
\usepackage{subcaption}
\usepackage{sidecap}
\usepackage[dvipsnames]{xcolor}
\graphicspath{{.}{./figures-main/}}
\usepackage{nameref,zref-xr}
\zxrsetup{toltxlabel}
\usepackage[colorlinks,linkcolor=blue,citecolor=blue,anchorcolor=blue]{hyperref}
\usepackage{url}
\usepackage{cleveref}
\begin{document}
\title{Strain-induced structural transitions in (111)-oriented (LaMnO$_3$)$_{2n}|$(SrMnO$_3$)$_n$ superlattices}

\author{Imran Ahamed}
\affiliation{Institute of Physics, Nicolaus Copernicus University, 87-100 Toru\'n, Poland}
\author{Shivalika Sharma}
\affiliation{Institute of Physics, Nicolaus Copernicus University, 87-100 Toru\'n, Poland}
\author{Fabrizio Cossu}
\affiliation{School of Physics, Engineering and Technology, University of York, York YO10 5DD, UK}
\affiliation{Department of Physics, School of Natural and Computing Sciences, University of Aberdeen, Aberdeen, AB24 3UE, United Kingdom}
\affiliation{Department of Physics and Institute of Quantum Convergence and Technology, Kangwon National University, Chuncheon, 24341, Republic of Korea}
\author{Igor {Di Marco}}
\email{igor.dimarco@physics.uu.se}
\email{igor.dimarco@umk.pl}
\affiliation{Institute of Physics, Nicolaus Copernicus University, 87-100 Toru\'n, Poland}\affiliation{Department of Physics and Astronomy, Uppsala University, Uppsala 751 20, Sweden}
\date{\today}

\begin{abstract}
By means of first-principles electronic structure calculations, we hereby investigate the structural transitions induced by epitaxial strain in (111)-oriented (LaMnO$_3$)$_{2n}|$(SrMnO$_3$)$_n$ superlattices, with $n=2,4,6$. All superlattices in the explored range of strain are shown to prefer a half-metallic ferromagnetic order where the local magnetic moments are coupled to volume-breathing distortions. More in detail, our results reveal that thickness plays a crucial role in the response to epitaxial strain, which is particularly evident in the resulting tilt pattern of the oxygen octahedra. The thinnest superlattice, for $n=2$,  always adopts the $a^-a^-a^-$ tilt pattern and the competing $a^-a^-c^+$ tilt pattern can be stabilized as a metastable state only in presence of compressive strain.
Instead, the superlattice with $n=4$ favours the $a^-a^-c^+$ tilt pattern at equilibrium conditions, but the in-phase rotations around the third pseudocubic axis are so fragile that the $a^-a^-a^-$ pattern is recovered under a tiny amount of either compressive or tensile strain.
The superlattice with $n=6$ exhibits a more nuanced behaviour: compressive strain drives a transition from $a^-a^-c^+$ to $a^-a^-a^-$, whereas tensile strain preserves the $a^-a^-c^+$ tilt pattern and significantly accentuates the structural differences between the two inequivalent sublattices within this symmetry. In fact, the Jahn-Teller distortions are quenched in one of the sublattices, leading to enhanced volume-breathing distortions and corresponding enhanced charge and spin oscillations. This suggests that Hund's physics may be more relevant in this regime of tensile strain, maximizing the interplay between strong electronic correlations and structural effects.
\end{abstract}

\maketitle

\section{Introduction}
Perovskite-oxide heterostructures provide a uniquely versatile setting to engineer correlated phases at the nanoscale, due to the interplay of charge, spin, orbital, and lattice degrees of freedom~\cite{Imada1998,Dagotto2005,hellman2017rmp,bhattacharya-AnnuRevMR2014}.
Among the available tuning parameters, epitaxial strain stands out as a one of the most practical control knobs to reshape bond lengths and angles, thus changing the hierarchy of competing structural instabilities. This mechanism is particularly consequential for manganites, where small changes in electronic and structural degrees of freedom may change the balance between double-exchange ferromagnetism and superexchange antiferromagnetism~\cite{PhysRevB_Zhao,Sen2007,lee_jh-PRB.88.174426}.
At the same time, strain can enhance or quench Jahn--Teller (JT) and volume-breathing (VB) distortions, eventually inducing carrier localization and orbital order~\cite{Schmitt2020,zhang2012,khomskii-JETP2016,Khomskii_2022,priyanka_preprint}.

Heterostructures and superlattices composed of LaMnO$_3$ (LMO) and SrMnO$_3$ (SMO) are a prototypical platform in which these couplings can be explored in a controlled, charge-balanced setting.
Bulk LMO is an A-type antiferromagnetic (AFM) Mott insulator with strong cooperative JT distortions~\cite{Schmitt2020,KHOMSKII202498}.
Bulk SMO, in contrast, is close to the cubic perovskite limit and behaves as a G-type AFM band insulator~\cite{KHOMSKII202498,Zhu2020SrMnO3}.
When combined in atomically precise superlattices, interfacial charge transfer and mixed Mn$^{3+}$/Mn$^{4+}$ valence promote emergent ferromagnetism and metallicity that are absent in either parent compound~\cite{smadici2007,bhattacharya2008,adamo2008,may2008,Nanda2008,nanda2009,Nanda2010,smadici2012,zhang2012,Nakao2015,keunecke2020}.
Crucially, these effects are inseparable from the oxygen-octahedra network, as octahedral connectivity couples rotation and tilt patterns between layers, and can strongly influence electronic transport and magnetism~\cite{zhou2020,tebano2008prl}.
While most LMO/SMO work has focused on the conventional (001) growth direction, the (111) orientation offers a qualitatively different landscape.
Along (111), the MnO$_6$ octahedra share three oxygens across the interface, enforcing a more coherent propagation of structural modes~\cite{Chakhalian2020} and potentially hosting correlated topological features~\cite{Chakhalian2020,xiao-ncomm2011,weng-PRB.92.195114,Chandra2017}.
In a previous first-principles study, we have shown that (111)-oriented (LMO)$_{12}|$(SMO)$_6$ superlattices can sustain robust half-metallic ferromagnetism across a wide thickness range~\cite{Cossu2022}. VB and JT distortions are tied to layer-resolved charge and spin oscillations, signalling an intrinsic coupling between lattice modes, charge and mixed-valence characteristics~\cite{Cossu2022}. More recently, we demonstrated that the same superlattices can display a mixed structural order, where distinct patterns of octahedral rotations and local MnO$_6$ distortions coexist and compete as a function of thickness~\cite{Cossu2024}.
Mixed order provides a natural mechanism for accommodating the geometric frustration and polar constraints inherent to the (111) stacking. 
It also offers a route to spatially patterned charge redistribution and moment modulation without introducing chemical disorder or charge doping. 

It is natural to ask how robust the aforementioned structural landscape is to epitaxial strain and if the latter could be exploited to selectively stabilize particular distortion patterns. Applying epitaxial strain in the (111) plane is expected to be very impactful, due to the mixing of shear-like and trigonal components, which can favour distinct combinations of rotation axes and JT/VB modes.
These questions are not only interesting from a fundamental perspective, but they can also guide experimental work in the field. 
Progress in growth and characterization has made (111)-oriented heterostructures of manganites achievable in experiment~\cite{Chakhalian2020,zhou2020,xu2023,wang2022pss,jansen2024prm}. Exploring the possibilities offered by the choice of suitable substrates and buffer layers is thus the next natural step.

In this study, we employ first-principles calculations to investigate the role of epitaxial strain on the structural properties of (111)-oriented (LMO)$_{2n}|$(SMO)$_n$ superlattices, for $n=2,4,6$. The total periodicity is chosen in multiple of six to accommodate the most favourable tilt patterns, while the largest thickness corresponds to the previously identified limit for the stability of the half-metallic ferromagnetic (FM) order~\cite{Cossu2022}.  
For each state, we investigate the ground-state magnetic order and the layer-resolved patterns of structural, magnetic and electronic properties. 
We found that thickness plays a crucial role in the response to epitaxial strain, which is particularly evident in the resulting tilt pattern of the oxygen octahedra. The complexity of the response is the highest for $n=6$, where a mixed structural order emerges and is amplified by tensile strain. Tracking the behavior of tilt patterns, van Vleck distortions, charges and local magnetic moments under epitaxial constraint provides a microscopic route to understand coupled lattice--charge--spin textures in (111)-oriented manganite superlattices.

The present paper is organized as follows. After this Introduction, section~\ref{sec:methods} is dedicated to the methodological details, including main equations and computational settings. The results of our study are presented in section~\ref{sec:results}, while our Conclusions are finally discussed in section~\ref{sec:conclusions}.


\section{Computational methods}\label{sec:methods}
Superlattices of chemical formula (LMO)$_{2n}|$(SMO)$_n$ with $n=2,4,6$ were studied by means of the density-functional theory (DFT) as implemented in the Vienna {\it ab-initio} Simulation Package (VASP)~\cite{vasp1,vasp2,vasp3}. The exchange-correlation functional was set to the Perdew–Burke–Ernzerhof (PBE) form of the generalised gradient approximation~\cite{gga}. The plane-wave kinetic energy cut-off was instead fixed to 500 eV. The projector augmented wave potentials~\cite{paw} used for Sr, La, Mn and O atoms included 4\textit{s}$^2$4\textit{p}$^6$5\textit{s}$^2$, 5\textit{s}$^2$5\textit{p}$^6$5\textit{d}$^1$6\textit{s}$^2$, 3\textit{s}$^2$3\textit{p}$^6$4\textit{s}$^2$3\textit{d}$^5$ and 2\textit{s}$^2$2\textit{p}$^4$ valence electrons, respectively. The structural optimisation of the superlattices with $n=2,4,6$ was performed by using a $\Gamma$-centered $\mathbf{k}$-mesh of $7 \times 4 \times 3$, $7 \times 4 \times 2$, and $7 \times 4 \times 1$, respectively.
The structural optimisations were performed with a tolerance of $1 \times 10^{-6}$ eV for the electronic loop with a Gaussian smearing of 10 meV and forces were converged up to $3 \times 10^{-5}$ eV\AA$^{-1}$. The self-consistent field calculations  to determine the final electronic and magnetic properties were performed with a convergence criterion of $1 \times 10^{-7}$ eV using the tetrahedron method. The rotationally invariant DFT+\textit{U} approach~\cite{Liechtenstein1995} is used to describe the localized 3\textit{d} orbitals \cite{Mellan2015} of the Mn-atoms, with Hubbard parameter, \textit{U} = 3.8 eV and Hund parameter, \textit{J} = 1.0 eV~\cite{Nanda2008,Nanda2010,Cossu2022,Cossu2024}. As pointed out in previous literature~\cite{Mellan2015,Schmitt2020, Cossu2022}, these parameters give a very good description of magnetic and structural properties, but tend to underestimate the bandgap in insulating solutions, which is not a problem for the scope of our work. Most importantly, these parameters offer a very good agreement~\cite{Cossu2022} with results obtained with the parameter-free meta-GGA functional SCAN~\cite{scan1}, which strengthens the validity of the results reported here.

For each structure, the in-plane lattice parameters are changed to model a biaxial strain in the [111]-plane of the superlattices.
The precise value of the strain is calculated with respect to the optimised lattice constant of 3.86~\AA\ obtained for the superlattice with $n=6$, as discussed in our previous study~\cite{Cossu2022,Cossu2024}. Here we confirm that this optimized value is unchanged for $n=2,4$ as well.
A realistic strain window going from $-3\%$ (compressive) to $+3\%$ (tensile) is considered, as representative of common perovskite substrates such as LaAlO$_3$ ($-1.8\%$), NdGaO$_3$ (about $0\%$), SrTiO$_3$ ($+1.2\%$), DyScO$_3$ ($+2.2\%$) and GdScO$_3$ ($+2.9\%$)~\cite{biegalski2005thermal,bark2011,Schmidbauerkd5054}. Under the constraints imposed by the applied strain, full ionic relaxation is performed for each configuration.

The distributions of locally-projected charges and magnetic moments across the superlattice layers are computed according to Bader theory~\cite{Bader1985,Henkelman2006,Henkelman2007,Henkelman2009,Yu2011}. The van Vleck distortions comprising VB distortion ($Q_1$) and JT distortions ($Q_2$ and $Q_3$) are defined as~\cite{Cossu2022,Schmitt2020,VanVleck1939}:
\begin{equation}
    Q_1 = (\Delta x + \Delta y + \Delta z)/\sqrt{(3)}
\end{equation}
\begin{equation}
    Q_2 = (\Delta x - \Delta y)/\sqrt{(2)}
\end{equation}
\begin{equation}
     Q_3 = (-\Delta x - \Delta y + 2\Delta z)/\sqrt{(6)}
\end{equation}
where $\Delta x$, $\Delta y$ and $\Delta z$ are the variations of the octahedral lengths $x$, $y$ and $z$ with respect to their average values.

\section{Results}\label{sec:results}
\begin{figure}[t]
\centering
\includegraphics[trim= 0.5cm 1cm 0.2cm 1.0cm, clip, width=1\linewidth]{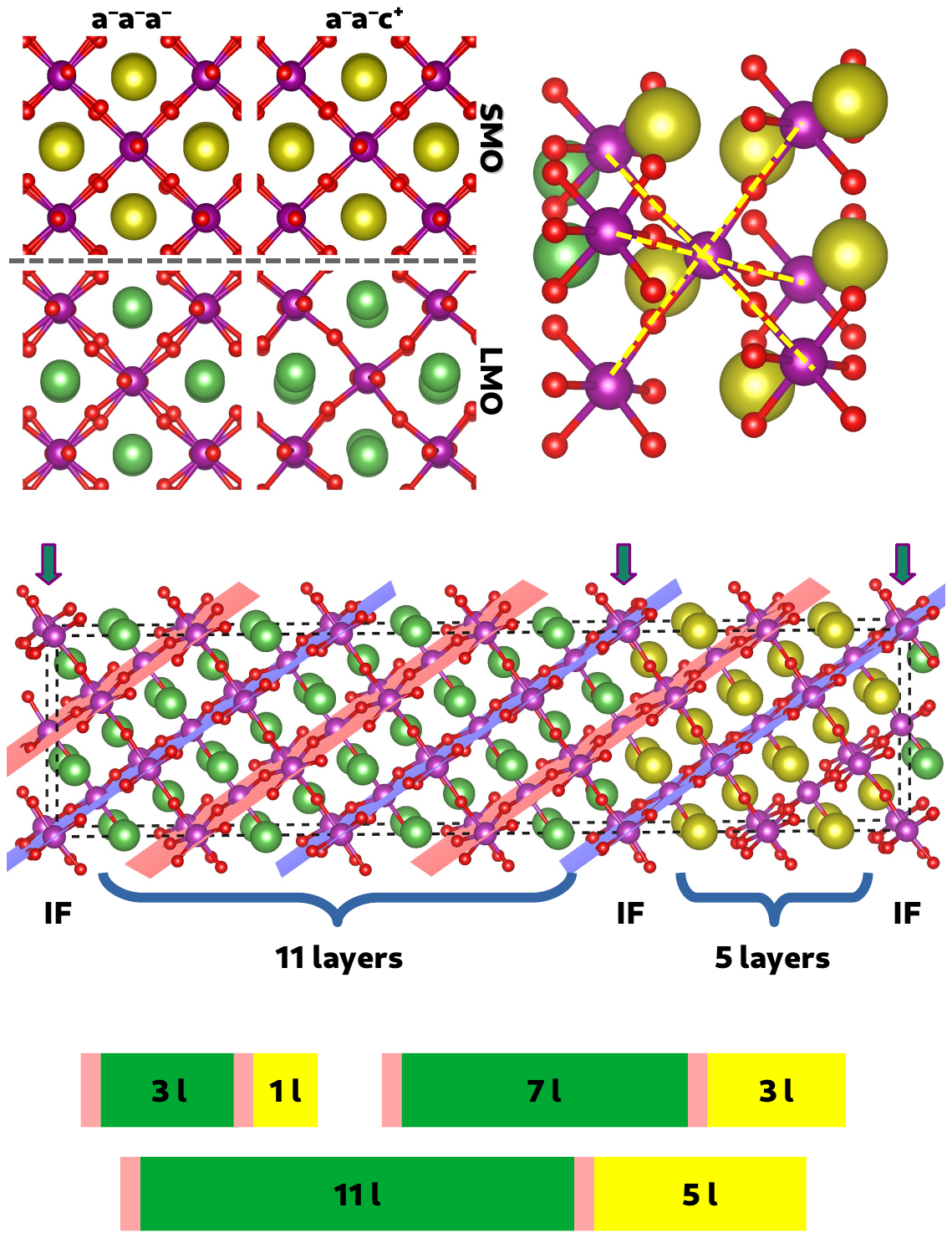}
\caption{Illustration of the structural characteristics of the (111)-oriented (LMO)$_{2n}|$(SMO)$_n$ superlattices. La, Sr, Mn and O ions are depicted as green, yellow, purple and red spheres, respectively. Top left panels: $a^-a^-a^-$ and $a^-a^-c^+$ tilting patterns as they relax in the SMO and LMO regions. Top right panel: local environment around a Mn atom, showing six connections (dashed yellow lines) with three Mn atoms towards an adjacent layer and three Mn atoms towards the opposite adjacent layer. Mid panel: side view of the $n=6$ superlattice, with illustration of the two Mn sublattices $S_e$ and $S_o$, shown in red and blue planes respectively; the interfacial (IF) layers are indicated by the blue arrows, while the layer count of the LMO and SMO regions is also emphasized. The $S_e$ and $S_o$ planes repeat in alternating patterns across the Mn-layers for the A-type AFM solution. Bottom panel: sketch of the layer count in the component regions for all the superlattices investigated in this study.}
\label{Fig1}
\end{figure}

For the investigation of the (111)-oriented (LMO)$_{2n}|$(SMO)$_n$ superlattices, two tilt patterns of the oxygen octahedra were considered, in agreement with our previous works~\cite{Cossu2022,Cossu2024}. These patterns can be labeled as $a^-a^-a^-$ and $a^-a^-c^+$ in Glazer's notation~\cite{glazer-AC:B1972,glazer-AC:A1975}, and are visualized in the top panels of Figure~\ref{Fig1}. The resulting structures belong to the space groups $R\bar{3}c$ and $Pnma$, respectively.
For each tilt pattern, four different magnetic orders were considered, namely FM and AFM of type A, C and G~\cite{KHOMSKII202498}. The ground state was found to be always FM, independently on $n$, tilt pattern and applied strain, at least within the range we explored. The FM ground state is also characterized by a half-metallic character that persists across all the layers of the superlattice. The consequences of these physical characteristics have already been analyzed in our first article on the subject and we redirect the reader to it for a more detailed analysis of these themes~\cite{Cossu2022}. Instead here we are going to focus on the effect of strain on the FM ground state, with a particular attention to the structural properties.

A schematic view of the investigated superlattices for $n=2,4,6$ is provided in the bottom panels of Figure~\ref{Fig1}, showing the LMO (green) and SMO (yellow) regions as well as interfacial (pink) layers. The full depiction of the ionic structure is instead shown in the middle panel, specifically for $n=6$. With the (111) geometry in mind, we proceed to analyze the structural order with respect to the thickness. For $n=2$, the ground state without applied strain adopts the $a^-a^-a^-$ tilt pattern, while the  $a^-a^-c^+$ tilt pattern cannot even be stabilized as a metastable state~\cite{Cossu2024}. While  the $a^-a^-a^-$ tilt pattern remains the ground state also under strain, a solution with the $a^-a^-c^+$ octahedral tilt pattern emerges for compressive strain, albeit as metastable.
Things are different for $n=4$, where we obtain that the ground state without applied strain adopts the $a^-a^-c^+$ tilt pattern from the outset. The $a^-a^-a^-$ tilt pattern can still be obtained as a metastable state, as we discussed in our prior study~\cite{Cossu2024}.
Interestingly, the $a^-a^-c^+$ tilt pattern is unstable toward both tensile and compressive strain. A structural transition to the $a^-a^-a^-$ tilt pattern happens already at a strain of $\pm 0.5\%$. A closer inspection of the $a^-a^-c^+$ tilt pattern reveals the reasons why this order is so unstable. The in-phase rotation around the third pseudocubic axis is present only in the innermost layers of the LMO region, and is so small that the system is close to a $a^-a^-c^0$ tilt pattern. This fragile order is disrupted as soon as a tiny epitaxial strain is applied. 
The situation observed for $n=6$ is instead more conventional. As already reported in our previous study~\cite{Cossu2022}, the ground state adopts the $a^-a^-c^+$ tilt pattern, but a small compressive strain of $-1.5\%$ is sufficient to induce a structural transition to the $a^-a^-a^-$ tilt pattern. For tensile strain, conversely, the structural order remains identical. Note that a similar structural transition has been observed in heterostructures of CaTiO$_3$ and LaNiO$_3$~\cite{kim2020}. Having established the obtained structural order under strain, we are now going to analyze the details of the structural distortions and magnetic properties separately for each thickness $n$. For simplicity, we will refer to the local magnetic moments as ``spins'' in the following.

\begin{figure}[t]
\centering
\includegraphics[clip, width=1\linewidth]{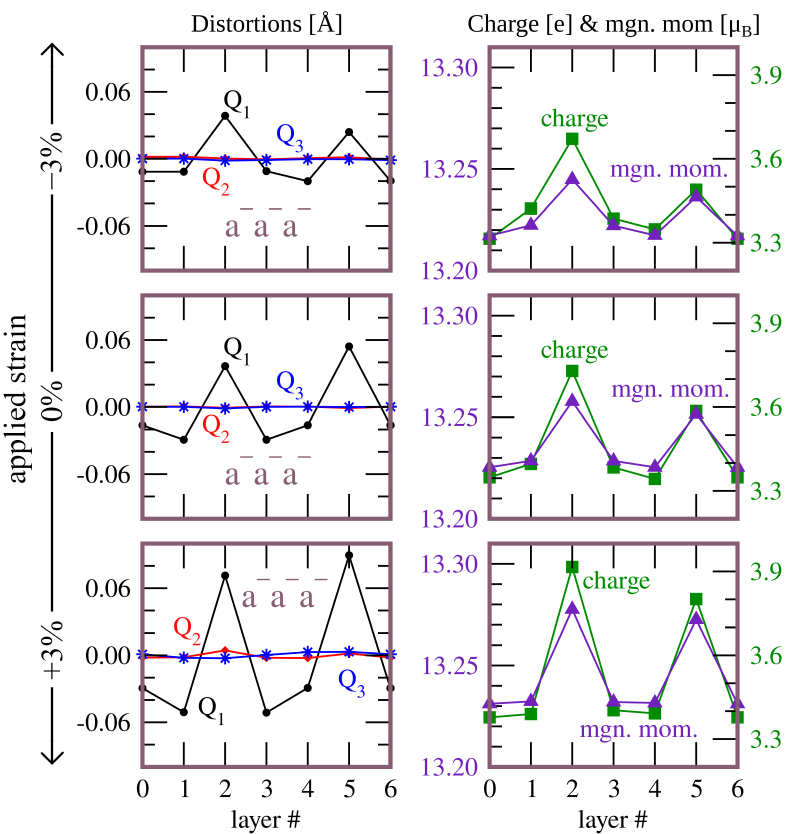}
\caption{Layer-resolved van Vleck distortions (left panels) as well as charge and spin distributions (right panels) for (111)-oriented (LMO)$_{2n}|$(SMO)$_n$ superlattices with $n=2$ in the FM ground state. Results obtained for compressive strain ($-3\%$), without strain ($0\%$) and for tensile strain ($+3\%$) are shown, respectively from top to bottom. The tilt pattern obtained for each strain is shown in the leftmost panels and is also indicated by a different color of the quadrants: aquamarine for $a^-a^-c^+$ and light brown for $a^-a^-a^-$. Finally, the $S_o$ and $S_e$ sublattices are identical by symmetry for the $a^-a^-a^-$ tilt pattern ($R\bar{3}c$ space group).}
\label{Fig2}
\end{figure}
The layer-resolved van Vleck distortions obtained for the ground state of the (111)-oriented (LMO)$_{2n}|$(SMO)$_n$ superlattices with $n=2$ are shown in the left panels of Figure~\ref{Fig2}, for compressive strain ($-3\%$), without strain ($0\%$) and for tensile strain ($+3\%$), respectively. We observe that the JT distortions $Q_2$ and $Q_3$ are always quenched for this thickness. The former is mainly quenched because of the symmetry of the $a^-a^-a^-$ tilt pattern, corresponding to the $R\bar{3}c$ space group. Previous studies have demonstrated that $Q_2$ in parent LMO does not have an electronic origin but arises due to a steric effect that affects the electron-lattice coupling, and thus depends crucially on the tilt pattern~\cite{varignon19prb,varignon19prr}.
In this case, however, also $Q_3$ is quenched, and this situation persists for all investigated strains. The overall trend observed with respect to the applied strain is trivial. For increasing in-plane lattice spacing, there is an increase of the VB distortions, which also amplifies the relative differences between the component regions of the superlattice. Layer-resolved charge and spin distributions, shown in the right panels of Figure~\ref{Fig2}, mirror this trend, showing larger charge differences and stronger magnetic moments for tensile strain. Conversely, under compressive strain, a more uniform behavior arises in VB distortions as well as charge and spin distributions. Overall, the superlattice with $n=2$ behaves similary to a mixed alloy of analogous composition, namely La$_{2/3}$Sr$_{1/3}$MnO$_3$.

\begin{figure}[b]
\centering
\includegraphics[clip, width=1\linewidth]{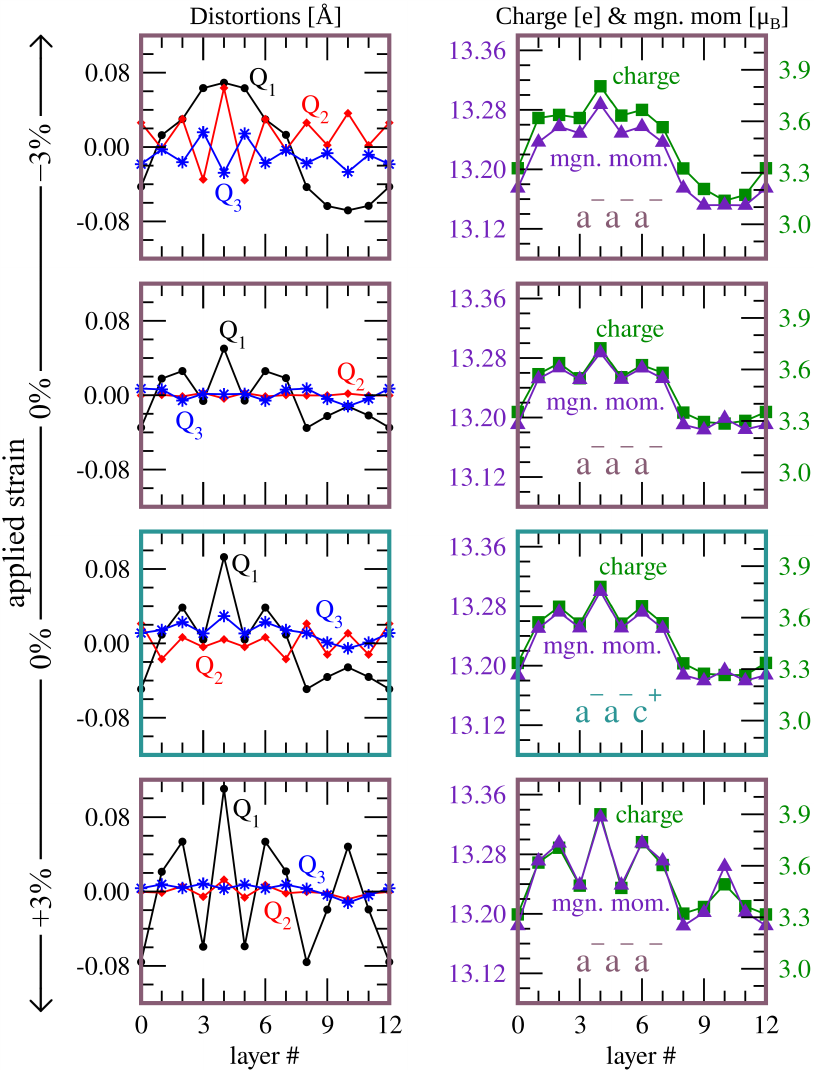}
\caption{Layer-resolved van Vleck distortions (left panels) as well as charge and spin distributions (right panels) for (111)-oriented (LMO)$_{2n}|$(SMO)$_n$ superlattices with $n=4$ in the FM ground state. Results obtained for compressive strain ($-3\%$), without strain ($0\%$) and for tensile strain ($+3\%$) are shown, respectively from top to bottom. The tilt pattern obtained for each strain in shown in the rightmost panels and is also indicated by a different color of the quadrants: aquamarine for $a^-a^-c^+$ and light brown for $a^-a^-a^-$. Finally, the $S_o$ and $S_e$ sublattices are identical by symmetry for the $a^-a^-a^-$ tilt pattern ($R\bar{3}c$ space group). For the $a^-a^-c^+$ tilt pattern, the small in-phase rotation around the third pseudocubic axis makes $S_o$ and $S_e$ quasidegenerate, so only one of them is reported in the plots.}
\label{Fig3}
\end{figure}
\begin{figure*}[t]
\centering
\includegraphics[trim= 0cm 0cm 0cm 0cm, clip, width=1.0\linewidth]{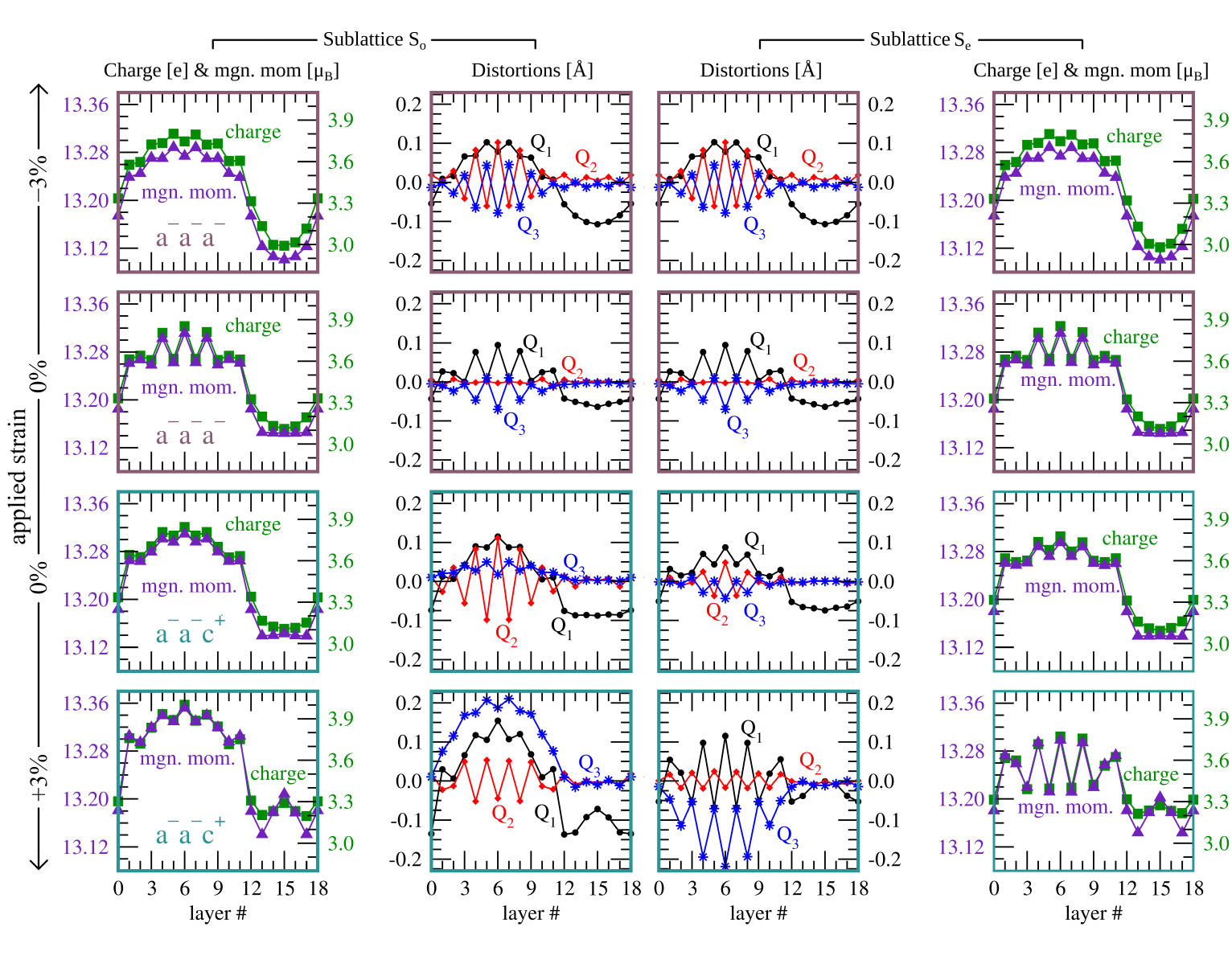}
\caption{Layer-resolved van Vleck distortions (central columns) and charge/spin distributions (left and right columns) for (111)-oriented (LMO)$_{2n}|$(SMO)$_n$ superlattices with $n=6$ in the FM ground state. Results are reported separately for the sublattice $S_o$ (left side) and for the sublattice $S_e$ (right side). Results obtained for compressive strain ($-3\%$), without strain ($0\%$) and for tensile strain ($+3\%$) are shown, respectively from top to bottom. The tilt pattern obtained for each strain in shown in the leftmost panels and is also indicated by a different color of the quadrants: aquamarine for $a^-a^-c^+$ and light brown for $a^-a^-a^-$. }
\label{Fig4}
\end{figure*}
The physical characteristics of the (LMO)$_{2n}|$(SMO)$_n$ superlattices with $n=4$ are much more complex. As remarked above, the system exhibits a ground state with the $a^-a^-c^+$ tilt pattern, but as soon as a small strain is applied, either compressive or tensile, a transition to the $a^-a^-a^-$ tilt pattern is obtained. As illustrated in middle panel of Figure~\ref{Fig1}, for the $a^-a^-c^+$ tilt pattern, corresponding to the $Pnma$ space group, the Mn sublattice can divide into two distinct sublattices, labeled as odd ($S_o$) and even ($S_e$), for convenience. However, the fact that our structural order is very close to the $a^-a^-c^0$ tilt pattern makes those sublattices quasi-degenerate in our case. They remain degenerate by symmetry for the $a^-a^-a^-$ tilt pattern. The layer-resolved van Vleck distortions as well as charge and spin distributions obtained for compressive strain ($-3\%$), without strain ($0\%$) and for tensile strain ($3\%$), are shown in Figure~\ref{Fig3}. When comparing the two tilt patterns obtained without strain, we observe that their properties are extremely similar, with the van Vleck distortions dominated by VB modes that are coupled to charge and spin oscillations reaching their maximum/minimum inside the LMO/SMO region. A noticeable difference is that $Q_2$ and $Q_3$ are larger in the $a^-a^-c^+$ tilt pattern, since the former is no longer quenched by symmetry (see discussion above). The behavior observed for tensile strain is similar to the one observed for $n=2$, i.e. the VB distortions become more marked, reflecting a stronger localization and larger spins. For compressive strain, the changes are more interesting. We first notice that $Q_2$ is no longer suppressed but becomes almost as large as $Q_1$. This means that either the structural changes due to compression have modified the structure of the superlattice so much to make the steric-driven $Q_2$ possible or the charge redistribution has made electronic JT possible. Considering that electronic JT usually responds negatively to a compression~\cite{lee_jh-PRB.88.174426,Banerjee2019prb}, we believe the first hypothesis to be more likely. The VB distortions do not exhibit marked oscillations any longer, but have a smooth profile, which is again perfectly mirrored by charge and spin distributions.

We then move to the most interesting case of $n=6$, where we analyze the effect of strain on structural distortions, charge and spin distributions, as well as on the mixed order between the $S_o$ and $S_e$ sublattices within the $a^-a^-c^+$ tilting pattern. Starting from the unstrained configuration~\cite{Cossu2024}, compressive strain ($-3\%$) drives a transition to the $a^-a^-a^-$ tilt system. In this symmetry setting, the mixed order is symmetry-forbidden.
Concurrently, the VB distortions follow an oscillatory trend similar to that observed under electron doping~\cite{Cossu2022}: the highest peak shifts from the central layer to the two adjacent layers, and the overall oscillation amplitude is reduced.
The JT distortions, which project onto the two-dimensional $E_g$ irreducible representation through the $Q_2$ and $Q_3$ modes, exhibit a distinctive behavior within the LMO region. In the unstrained case both $Q_2$ and $Q_3$ are very small, so that their relative orientation within the $E_g$ manifold is not well defined. Under compressive strain, however, a robust $E_g$ distortion develops (see topmost panels of Figure~\ref{Fig4}), with both $Q_2$ and $Q_3$ acquiring finite amplitude. The layer-resolved analysis shows that the two components maintain an approximately fixed internal ratio (in our case $Q_3 \simeq -Q_2$), indicating that strain stabilizes a specific direction within the $E_g$ manifold rather than activating two independent distortion channels. The corresponding total JT amplitude, $Q = \sqrt{Q_2^2 + Q_3^2}$,
remains moderate and follows the layer-by-layer modulation of this staggered $E_g$ order parameter.
The evolution of the VB distortions is closely mirrored by the layer-resolved charges and magnetic moments. In particular, we note the presence of a dip in place of a peak in the central layer of the LMO region and a shallow minimum within the SMO region, consistent with a net charge redistribution from the Mn centers toward the O ligands.

If compressive strain stabilizes the $a^-a^-a^-$ tilting pattern, tensile strain ($+3\%$) instead maintains the $a^-a^-c^+$ configuration of the ground state. In this phase, the mixed order is not suppressed; rather, it becomes more pronounced. The system separates into two symmetry-inequivalent Mn--O--Mn chains, each characterized by a distinct layer-resolved pattern of $Q_1$, $Q_2$, $Q_3$, charge, and spin. This is in stark contrast with the $a^-a^-a^-$ phase, where the two sublattices are equivalent by symmetry.
Already at zero strain within the $a^-a^-c^+$ structure, the two sublattices display qualitatively different JT behaviour: one exhibits large oscillations of $Q_2$ accompanied by small $Q_3$, while the other shows only weak $Q_2$ modulations. Under tensile strain this differentiation is enhanced, as shown in the bottom panels of Figure~\ref{Fig4}. In the first sublattice, $Q_2$ retains large oscillations and $Q_3$ develops a sizable, smoothly-variable component. In the second sublattice, $Q_2$ remains weakly modulated, whereas $Q_3$ acquires pronounced oscillations.
The VB mode $Q_1$ follows the same sublattice-selective trend: it is relatively smooth in the first sublattice and strongly modulated in the second, with tensile strain further amplifying this contrast. These results indicate that tensile strain reinforces the inequivalence between the two sublattices, enhancing the mixed structural, electronic, and magnetic order characteristic of the $a^-a^-c^+$ phase.
Notably, the shallow, smooth minimum for atomic volumes, charges and spins in the SMO region that is found under compressive strain becomes a roughly constant value in absence of strain and then is replaced by a central peak for tensile strain. Contextually, in the LMO region, a slight increase of the charge and magnetic moment characterizes sublattice $S_o$, whereas sublattice $S_e$ features a slight charge depletion (notice the oscillations of charge and magnetic moments).

\section{Conclusions}\label{sec:conclusions}
We have investigated how epitaxial strain affects the structural properties, and the magnetic properties coupled to them, in (111)-oriented (LaMnO$_3$)$_{2n}|$(SrMnO$_3$)$_n$ superlattices, with $n=2,4,6$. We have found that the response to applied strain depends crucially on the thickness of the superlattice. While the thinnest superlattice, for $n=2$, shows a trivial response to both compressive and tensile strain, a more complex behavior is observed for larger thickness. For $n=4$, we observe that strain disrupts the fragile order of the $a^-a^-c^+$ tilt pattern by inducing a transition to the $a^-a^-a^-$ tilt pattern for either compressive or tensile strain. For tensile strain, marked charge and spin oscillations can be observed, driven by dominant VB distortions. For compressive strain, JT distortions are shown to arise even in the $a^-a^-a^-$ tilt pattern, where they are expected to be quenched.
For $n=6$, instead, compressive and tensile strain are found to stabilize the $a^-a^-a^-$ tilt pattern and the $a^-a^-c^+$ tilt pattern, respectively. For the latter, tensile strain is found to enhance the mixed structural order observed in the sublattice differentiation. Thus, the choice of adequate substrates and capping layers seems a viable strategy to enhance the mixed structural order in experimental settings. State-of-the-art diffraction and microscopy can be used to resolve subtle oxygen-octahedra distortions and symmetry changes in buried layers,
providing a direct route to test strain-stabilized mixed-order scenarios~\cite{jansen2024prm,zhou2020}.

\vspace{1cm}
\section*{Acknowledgments}
We are thankful to I.\ E.\ Brumboiu, B.\ Sanyal and V.\ K.\ Lazarov for valuable discussions.
We acknowledge Polish high-performance computing infrastructure PLGrid for awarding this project access to the LUMI supercomputer, owned by the EuroHPC Joint Undertaking, hosted by CSC (Finland) and the LUMI consortium through PLL/2023/04/016450. Additional computational work was performed on resources provided by the National Academic Infrastructure for Supercomputing in Sweden (NAISS), partially funded by the Swedish Research Council through Grant Agreement No. 2022-06725. 
This research is part of the project No. 2022/45/P/ST3/04247 co-funded by the National Science Centre of Poland and the European Union's Horizon 2020 research and innovation programme under the Marie Sk{\l}odowska-Curie grant agreement no. 945339.
F.\ C. acknowledges financial support from the National Research Foundation (NRF) funded by the Ministry of Science of Korea (Grant No.\ 2022R1I1A1A01071974). 
The present project was also supported by the STINT Mobility Grant for Internationalization (Grant No. MG2022-9386).

\end{document}